\documentclass[lettersize,journal]{IEEEtran}
\usepackage{amsmath,amsfonts}
\usepackage{algorithmic}
\usepackage{algorithm}
\usepackage{placeins} % Add this to your preamble
\usepackage{array}
\usepackage{colortbl}
\usepackage[caption=false,font=normalsize,labelfont=sf,textfont=sf]{subfig}
\usepackage{textcomp}
\usepackage{tabularx}
\usepackage{stfloats}
\usepackage{url}
\usepackage{verbatim}
\usepackage{graphicx}
\usepackage{cite}
\usepackage{breqn}
\usepackage{bm}
\usepackage{multirow}
\usepackage{algorithm}
\usepackage{algorithmic}
\usepackage{tikz}
\usepackage{xcolor}
\newcommand{\mycolor}[2]{%
  \ifthenelse{\equal{#1}{black}}{\textcolor{black}{#2}}{\textcolor{#1}{#2}}%
}
\begin{document}

\title{Personalized Speech Emotion Recognition in Human-Robot Interaction using Vision Transformers}

\author{Ruchik Mishra,~\IEEEmembership{Student Member, IEEE}, 
        Andrew Frye, 
        Madan M. Rayguru, 
        Dan O. Popa, ~\IEEEmembership{Senior Member, IEEE}%
% \thanks{ Manuscript received: October 22, 2024; Revised: January 19, 2025; Accepted: March 1, 2025.}        
% \thanks{This paper was recommended for publication by Editor Angelika Peer upon evaluation of the Associate Editor and Reviewers' comments.}
\thanks{This project was supported in part by the National Institutes of Health (NIH) and the National Science Foundation (NSF) through the Smart and Connected Health (SCH) grant \#1838808, and in part through the EPSCoR grant \#1849213. Authors are with the Louisville Automation and Robotics Research Institute (LARRI), at the University of Louisville, KY, USA.}%
% \thanks{Madan Mohan Rayguru is with the College of Business, University of Louisville, KY.}%
% \thanks{Digital Object Identifier (DOI): see top of this page.}
}
% \markboth{IEEE Robotics and Automation Letters. Preprint Version. March, 2025}
% {FirstAuthorSurname \MakeLowercase{\textit{et al.}}: ShortTitle}
% \markboth{}%


\maketitle
\begin{abstract}
% Emotions are an essential element in verbal communication and so understanding affect of individuals during a human-robot interaction (HRI) becomes imperative. This paper focuses on Speech Emotion Recognition (SER) within an HRI context using vision transformer models. We evaluate both ViT (Vision transformer) and BEiT (BERT Pre-Training of Image Transformers) based pipelines separately to generalize over individual speech characteristics. We train on a few audio samples from each participant interacting with a NAO robot in a pseudo-naturalistic conversation. Then, we test our ensemble models on an unseen set of sentences but having the same emotional speech characteristics as the individual. In the results, we show that fine-tuning vision transformers on benchmark datasets and/or ensembling gets us the highest classification accuracies per individual when it comes to identifying four primary emotions from their speech: neutral, happy, sad, and angry as compared to fine-tuning vanilla-ViTs or BEiTs. \\
Emotions are an essential element in human verbal communication, therefore it is important to understand individuals' affect during human-robot interaction (HRI). This paper investigates the application of vision transformer models, namely ViT (Vision Transformers) and BEiT (Bidirectional Encoder Representations from Pre-Training of Image Transformers) pipelines for Speech Emotion Recognition (SER) in HRI. The focus is to generalize the SER models for individual speech characteristics by fine-tuning these models on benchmark datasets and exploiting ensemble methods. For this purpose, we collected audio data from several human subjects having pseudo-naturalistic conversations with the NAO social robot. We then fine-tuned our ViT and BEiT-based models and tested these models on unseen speech samples from the participants in order to dentify four primary emotions from speech: neutral, happy, sad, and angry.  The results show that fine-tuning vision transformers on benchmark datasets and then using either these already fine-tuned models or ensembling ViT/BEiT models results in higher classification accuracies than fine-tuning vanilla-ViTs or BEiTs.
\end{abstract}

\section{Introduction}
\IEEEPARstart{T}{he} increasing integration of social robots across various sectors, from healthcare to customer service, underscores their potential to revolutionize human-machine interaction \cite{johanson2021improving, Mishra2022, Mishra2023, nakanishi2020continuous}. A crucial factor in their application success is the ability to perceive and respond appropriately to human emotions, facilitating meaningful and engaging interactions \cite{Mishra2022, kirby2010affective, 9384222,9508862}. In this context, Speech Emotion Recognition (SER) emerges as a critical field within human-computer interaction \cite{Ibrahim2024}. By enabling machines to understand and respond to the emotional nuances embedded in human speech (affective speech), SER can transform our interactions with technology, fostering more natural and empathetic communication \cite{Ibrahim2024}. When social robots can accurately interpret affective speech, they can adapt their behavior and responses, leading to more personalized and impactful human interactions \cite{Mishra2022}. This emotional connection ability holds tremendous potential for enhancing the effectiveness and acceptance of social robots in various real-world applications.\\
The importance of affective speech in human-robot interaction (HRI) lies in its ability to enhance the robot's social intelligence and facilitate natural communication \cite{Ibrahim2024, Pepino2021}. Emotions play a fundamental role in human interactions. By understanding and responding to affective cues, robots can build trust, rapport, and cooperation with their human counterparts \cite{Bott2024}. Affective speech recognition capability enables social robots to accurately perceive the emotional state of the user, allowing them to tailor their responses and provide appropriate support or feedback \cite{Chang2022}.\\
The area of Speech Emotion Recognition (SER) has witnessed significant advancements over time, driven by the exploration of diverse feature extraction methods and suitable machine learning techniques. Early research focused on traditional approaches like Mel-frequency spectral coefficients (MFCCs) and prosodic features, laying the groundwork for subsequent extensions and improvements \cite{zhou2013speech, lalitha2015speech, rao2013emotion}. The emergence of deep learning further matured the area, with models like DNNs, RNNs, and CNNs demonstrating improved capabilities in capturing emotional nuances from speech \cite{han2014speech, lee2015high, satt2017efficient}. 
% The recent introduction of self-supervised learning (SSL) models, notably HuBERT and Wav2Vec2, has marked a turning point in SER research \cite{hsu2021hubert, pepino2021emotion}. 
% These models excel at learning rich representations from extensive unlabeled speech data, enabling effective fine-tuning for specific SER tasks and achieving state-of-the-art performance on benchmark datasets like IEMOCAP \cite{chen2023exploring,wagner2023dawn}. The ability of SSL models to leverage large-scale unlabeled data has unlocked new possibilities for SER, paving the way for more robust and accurate emotion recognition systems. We discussed more about these in the related works section.\\

% \begin{figure*}
%     \centering
%     \includegraphics[width=1.0\linewidth]{Pipeline.eps}
%     \caption{The experimental workflow showing our speech emotion recognition pipeline during human-robot interaction.}
%     \label{fig:pipeline}
% \end{figure*}
\begin{figure*}[h!]
    \centering
    \subfloat[Full Pipeline]{%
        \includegraphics[width=0.85\textwidth]{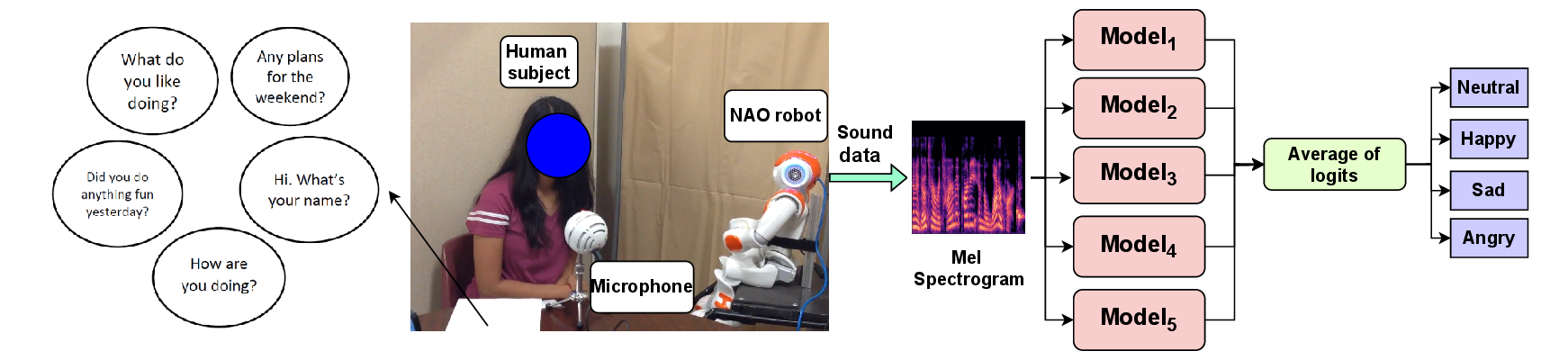}%
        \label{fig:pipeline}
    }
    
     \subfloat[Model$_{i}$, where $i \in \left\{1,2,3,4\right\}$. Transformer encoder here represents the ViT/BEiT encoder]{%
        \includegraphics[width=0.60\textwidth]{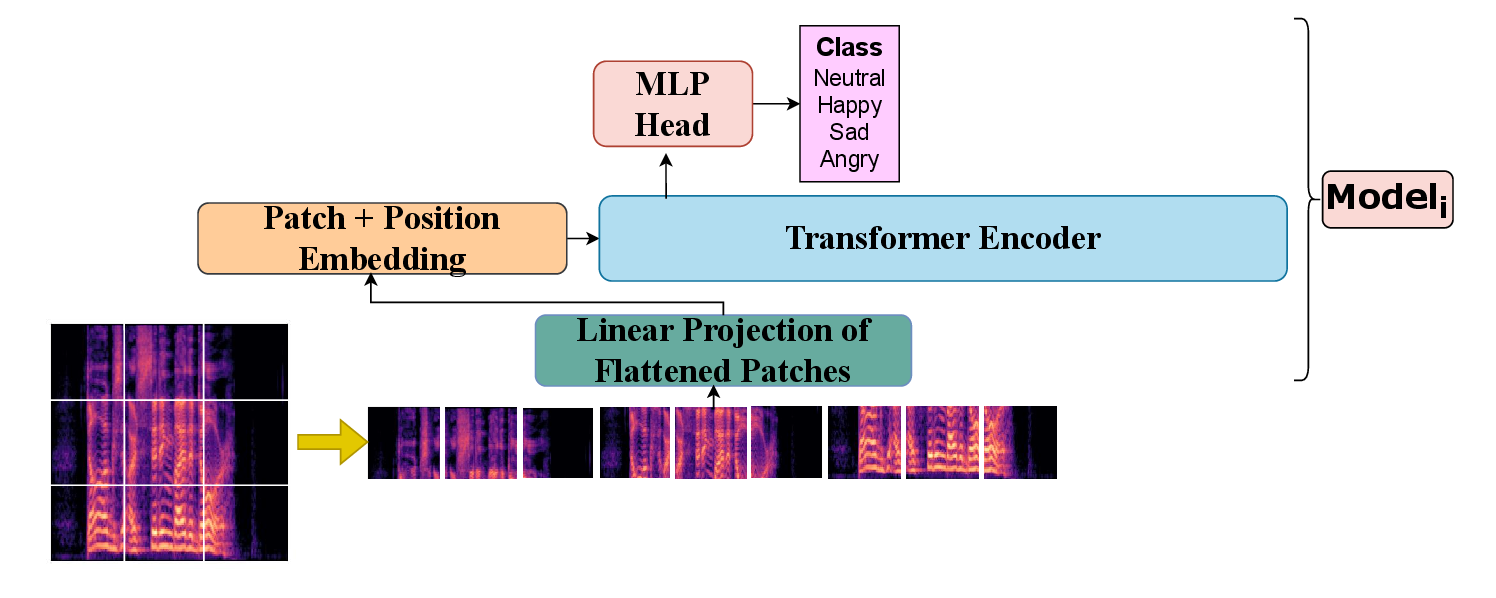}%
        \label{fig:ViT_model}
    }
    \caption{The two pipelines evaluated in this paper for speech emotion recognition.}
    \label{fig:transformer_pipline}
\end{figure*}

\textcolor{black}{Recent advancements in computer vision, particularly with the emergence of Vision Transformers (ViTs), have opened up new possibilities for leveraging visual data in SER \cite{khasgiwala2021vision}. This is evident from their superior performance as compared to other deep learning based approaches as shown in \cite{akinpelu2024enhanced}.}
% ViTs, originally designed for image classification tasks, have demonstrated exceptional performance in capturing spatial dependencies and global contextual information \cite{dosovitskiy2020image, bao2021beit}. Their ability to model long-range dependencies and extract high-level features makes them well-suited for analyzing visual representations of speech, such as spectrograms \cite{vaswani2017attention}. The application of ViTs in SER is still an emerging field, and several research directions warrant further exploration. 

% One area of interest is the development of lightweight ViT models specifically tailored for SER tasks. 
% These models should be computationally efficient to enable real-time emotion recognition in resource-constrained environments. Another avenue is to investigate the fusion of visual and acoustic features at different levels of the ViT architecture to optimize the integration of multi-modal information.

% In conclusion, the integration of Vision Transformers (ViTs) in Speech Emotion Recognition (SER) represents an interesting approach to advance the human-computer interaction. By leveraging the power of ViTs to analyze visual representations of speech, we can develop more accurate, robust, and adaptable SER systems. The ability to understand and respond to human emotions in real-time has the potential to transform various applications, including virtual assistants, mental health monitoring, and human-robot interaction.  

In this work, we evaluate vision transformer based models for speech emotion recognition. To the best of our knowledge, this work is one of the earliest in the literature to evaluate vision transformer based models for speech emotion recognition in pseudo-naturalistic verbal communications in HRI. This evaluation of ViT based models has been done for modeling the individual characteristics in SER. This means, given a set of audio clips from an individual with labelled emotions (here neutral, happy, sad, and angry), we can predict the speech emotion of that individual for a different set of sentences spoken during a one-to-one HRI. To support our claim, we collect data from human participants an engage them in a pseudo-naturalistic conversation with the robot (explained more in section \ref{data_acquisition}).\\
% \textcolor{red}{\textbf{Contributions:} The major contributions are outlined as follows:
% \begin{itemize}
%     \item This work is among the first/earliest to investigate vision transformer models (ViT and BEiT) for SER, in the context of pseudo-naturalistic Human-Robot Interaction (HRI).
%     \item We propose a novel approach to personalize SER models by fine-tuning them on benchmark datasets and leveraging ensemble techniques.
%     \item Our experiments demonstrate state-of-the-art performance on several benchmark datasets and showcase the superior accuracy of our approach in capturing individual speech emotion characteristics during HRI.
%     \item The findings of this study contribute to the development of more emotionally intelligent and personalized social robots capable of understanding and responding appropriately to human emotions.
% \end{itemize}}
This paper makes the following contributions:
\begin{itemize}
    \item This work is among the first to investigate vision transformer-based models (both ViT and BEiT) for SER in the context of pseudo-naturalistic verbal HRI.
    \item We show that personalization of SER models can be done by fine-tuning ViT and BEiT models on benchmark datasets and then further fine-tuning these on participant data and through ensembling the models. 
    \item \textcolor{black}{We compare our ViT and BEiT models with OpenAI/Whisper-base and ResNet-50 models.}
    \item \textcolor{black}{We recruited both native and non-native English speakers to include more diverse demographics for robustness.}
    \item Lastly, we also achieve state-of-the-art (SOTA) performance on the RAVDESS and TESS datasets by a full fine-tuning of the vision transformer-based models.
\end{itemize}
This paper has been arranged in the following manner: Section \ref{related_works} outlines the background literature supporting this work. Section \ref{methodology} descibes the methodology, which includes the data acquisition (Section \ref{data_acquisition}), description about mel-spectrograms (Section \ref{mel_spectrograms}), datasets used (Section \ref{datasets}),  and problem formulation (Section \ref{problem_formulation}). This is followed by Section \ref{results}, which discusses the results we obtained, and followed by the conclusion in Section \ref{conclusion}.
\vspace{-0.4cm}
\section{Related Works}\label{related_works}
% In the literature, mel spectrograms have been used for a variety of audio classification tasks \cite{zaman2023survey}. These classification tasks include music genre classification.  \\

The evolution of Speech Emotion Recognition (SER) has been marked by a continuous exploration of increasingly sophisticated techniques, each building upon the foundations laid by its predecessors. Early research in SER relied heavily on traditional approaches, such as Mel-frequency cepstral coefficients (MFCCs) and prosodic features \cite{zhou2013speech, lalitha2015speech, rao2013emotion}. MFCCs, derived from the human auditory system's response to sound, capture spectral characteristics crucial for distinguishing various speech sounds, while prosodic features like pitch, intensity, and duration provide insights into the emotional tone of speech. These handcrafted features, though valuable, often struggled to capture the subtle and complex interplay of acoustic cues that contribute to emotional expression.

The advent of deep learning revolutionized the field of SER, offering a powerful framework for automatically learning intricate patterns and representations from raw speech data. Deep Neural Networks (DNNs), with their multiple layers of interconnected nodes, enabled the extraction of high-level features that better captured the subtle nuances of emotional speech \cite{han2014speech}. Recurrent Neural Networks (RNNs), particularly Long Short-Term Memory (LSTM) networks, proved adept at modeling the temporal dynamics of speech, crucial for understanding the evolution of emotions over time \cite{lee2015high}. Convolutional Neural Networks (CNNs), originally designed for image processing, demonstrated their effectiveness in capturing local patterns and spatial dependencies in spectrograms, further enhancing SER performance \cite{satt2017efficient}.

The authors in \cite{lakomkin2018robustness} proposed to use CNN and RNN pipelines along with data augmentation techniques to improve the robustness of these models. This robustness was crucial for a human-robot interaction scenario with robot's ego noise. \textcolor{black}{The authors in \cite{mishra2024speech} also used a CNN plus BiLSTM hybrid model for the SER task using SAVEE and TESS datasets}. Further, the authors in \cite{chen2020two} proposed a machine learning pipeline for SER. Their approach involves using personalized and non-personalized features for SER. However, neither of these papers contributes to evaluating transformer-based architectures, which are currently SOTA in numerous fields of study \cite{vaswani2017attention}.
\vspace{-0.11cm}

A number of benchmark datasets have been developed for SER that capture speaker characteristics owing to the number of actors involved for generating the data. More information about these datasets have been discussed in Section \ref{datasets}. Owing to this large number of datasets, numerous approaches have been proposed in the literature. Even with transformer-based architectures, limited work has been shown in the SER literature. The authors in \cite{luna2021proposal} show the highest performance on the Ryerson Audio-Visual Database of Emotional Speech and Song (RAVDESS) \cite{livingstone2018ryerson} (described more in Section \ref{datasets}), using a pre-trained xlsr-Wav2Vec2.0 transformer. A more recent transformer-based approach includes the work by the authors in \cite{ibrahim2024does} where they used a Whisper-based speech emotion recognition. Other attention mechanism-based approaches for the RAVDESS dataset include \cite{chumachenko2022self}. For the Toronto emotional speech set (TESS) \cite{dupuis2010tess}, authors in \cite{akinpelu2024enhanced} tested the accuracies for SER tasks using a vision-transformer-based architecture. These transformers-based approaches have also been evaluated on the Crowd Sourced Emotional Multimodal Actors Dataset (CREMA-D) \cite{cao2014crema,chen2024vesper}. The authors in \cite{chen2024vesper} tested their approach called the improVed emotion-specific pre-trained encoder (Vesper) on benchmark datasets like Multimodal EmotionLines Dataset (MELD) and Interactive Emotional Dyadic Motion Capture (IEMOCAP) database in addition to the CREMA-D. Further, the authors in \cite{saliba2024layer} approach to use Acoustic Word Embeddings (AWEs) to push the classification accuracies on the Emotional Speech Database (ESD) and IEMOCAP. 

\textcolor{black}{For transformer based SER models, some recent works have made attempts to model personalised features of users like the authors in \cite{liu2025personalized}. Other approaches specific to vision transformers based approached for SER include the work by the authors in \cite{ong2024maxmvit} where they have used the strengths of Multi-Axis Vision Transformer (MaxViT) and the Improved Multiscale Vision
Transformer (MViTv2).} 
\vspace{-0.07cm}

However, the literature on SER and the datasets available have not been extensively leveraged to model speaker characteristics in a one-to-one human-robot situation using these SOTA transformer architectures.
\vspace{-0.32cm}
\section{Methodology}\label{methodology}
\subsection{Data Acquisition}\label{data_acquisition}
% Six neurotypical adults were recruited to participate in a human-robot interaction study to classify their speech into four primary emotions. This one-to-one HRI involves two-way communication between the robot and the participants. 
\textcolor{black}{Twelve neurotypical participants were recruited to participate in a human-robot interaction study to classify their speech into four primary emotions. Six of these participants were native English speakers. The other six were non-native English speakers. This was done to include more diverse demographics to examine SER using vision transformers based models. Among the participants, five were male and the rest were female. All the participants were either students or staff from the university aged between 18-59 years of age.}
Each participant asks pre-defined questions as shown in Figure \ref{fig:pipeline}. These questions had been used for our previous studies during HRI \cite{mishra2023social}. The following are the questions we asked the participants to ask the robot: 
\begin{itemize}
    \item Hi. What's your name?
    \item How are you doing? 
    \item Did you do anything fun yesterday?
    \item What do you like doing?
    \item Any plans for the weekend?
\end{itemize}
The robot responds with appropriate answers to those questions and asks those questions back to the participant. The participants' replies are not pre-defined. They were asked to reply to the robot's questions with short answers. For each of these question-and-answer pairs, each participant was asked to speak in an emotional tone depicting one of the four primary emotions, i.e., neutral, happy, sad, and angry. The voices of the participants were recorded during this pseudo-natural human-robot interaction where the questions that the participant asks were pre-defined but their answers weren't. \textcolor{black}{More information on personalization is shared in Algorithm \ref{alg:personalization}.}
\vspace{-0.4cm}
\subsection{Mel Spectrogram}\label{mel_spectrograms}
In this paper, since we are using vision based models, we convert the sound signals to 2D images. This is where we leverage the use of mel spectrograms. The mel spectrogram is used for better perception of sounds by humans. Considering $f$ as the normal frequency, the frequency on the mel scale ($m$) will be given by \cite{ancilin2021improved,abdul2022mel, zhang2019audio}: 
\begin{equation}\label{eq:mel}
\begin{split}
        m = 2595 \log_{10} \left(1+ \frac{f}{700}\right)
     &  = 1125\ln \left(1+ \frac{f}{700}\right)
    \end{split}
\end{equation}
As can be seen form equation \ref{eq:mel}, the mel scale is a logarithmic scale to convert the frequency of the sounds from Hz to mels. The audio signal first goes through a fast Fourier transform performed on overlapping signal segments. These frequencies are converted to the log scale and the amplitude is converted to decibels to make the color dimension as shown in Figure \ref{fig:pipeline}.
\vspace{-0.3cm}
\subsection{Datasets}\label{datasets}
\vspace{-0.5cm}
\begin{table}[h!]
\centering
\caption{Total number of data points for each emotion label for all datasets used}
\begin{tabular}{|c|c|c|c|c|c|}
    \hline
    \multicolumn{6}{|c|}{\textbf{Datasets}} \\
    \hline
    % \multirow{2}{*}{\textbf{Heading}} & \textbf{Column 1} & \textbf{Column 2} & \multicolumn{2}{c|}{\textbf{Column 3 and 4}} \\
    % \cline{2-5}
    \textbf{Emotion} & \textbf{RAVDESS} & \textbf{TESS} & \textbf{CREMA-D} & \textbf{ESD} & \textbf{MELD}\\
    \hline
    Neutral & 96 & 359 & 1087 & 3500 & 6527\\
    \hline
    Happy & 192 & 350 & 1271  & 3500 & 2416\\
    \hline
    Sad & 192 & 352 & 1271 & 3500 & 917\\
    \hline
    Angry & 192 &370  & 1271 & 3510 & 1560\\
    \hline
\end{tabular}
\label{table:your_table_label}
\end{table}
For fine-tuning our vision transformer-based models, we use four benchmark datasets from the literature. 
\begin{itemize}
    \item \textbf{RAVDESS \cite{livingstone2018ryerson}: } This dataset has 1440 files containing data from 24 actors making sixty trials each. These actors cover seven emotions: calm, happy, sad, angry, fearful, surprise, and disgust. All of these emotions are deliberately displayed in the speech characteristics of each of the actors by speaking two sets of sentences, each with these seven emotional traits.  
    \item \textbf{TESS \cite{dupuis2010tess}:} TESS contains data from two actresses aged 26 and 64 years. Each of the actresses speak pre-defined sentences in different ways so as to create a total of 2800 stimuli. These cover seven emotions: happiness, sadness, fear, pleasant surprise, anger, disgust, and neutral. 
    \item \textbf{CREMA-D \cite{cao2014crema}:} This dataset captures six different emotions: happy, sad, neutral, anger, disgust, and fear. These stimuli were created by 91 actors generating a total of 7442 clips. 
    \item \textbf{ESD \cite{zhou2021seen} :} This dataset captures the speakers' emotions for five emotional classes: neutral, happiness, anger, sadness, and surprise. These emotional stimuli were recorded by 20 speakers, 10 of whom were native English speakers. 
    \item \textbf{MELD \cite{poria2018meld} : } It is a multiparty multimodal dataset that captures speakers' emotions from the TV-series Friends. This dataset captures emotions in both continuous and discrete ways. Among the discrete emotions, it captures seven emotions: anger, disgust, sadness, joy, neutral, surprise, and fear. 
\end{itemize}
For all of these datasets, we have used only four emotion classes that are common between these four datasets, i.e., neutral, happiness, sadness, and anger. In addition to it, we used only ten actors for the ESD dataset who were native English speakers.  
% \begin{figure*}[h!]
%     \centering
%     \subfloat[ViT model]{%
%         \includegraphics[width=0.5\textwidth]{vision_transformers_mel_spect.eps}%
%         \label{fig:vit_pipeline}
%     }
%      \subfloat[BEiT model]{%
%         \includegraphics[width=0.5\textwidth]{BEiT_mel_spect.eps}%
%         \label{fig:beit_pipeline}
%     }
%     \caption{The two pipelines evaluated in this paper for speech emotion recognition.}
%     \label{fig:transformer_pipline}
% \end{figure*}
\vspace{-0.5cm}
\subsection{Problem Formulation and Proposed Pipeline}\label{problem_formulation}
% \begin{figure}[h!]
%     \centering
%     \includegraphics[width=1.05\linewidth]{vision_transformers_mel_spect.eps}
%     \caption{\textcolor{black}{Transformer Encoder represents ViT/BEiT encoder, the two pipelines evaluated in this paper.}}
%     \label{fig:ViT_BEiT_encoder}
% \end{figure}
% Initially, we used both ViT and BEiT models on individual datasets to evaluate these models on three classification metrics: accuracy, precision, and recall. For each data point 
% As discussed in the previous section, we use four datasets $\mathcal{X} = \left\{\textrm{RAVDESS}, \textrm{TESS}, \textrm{CREMA-D}, \textrm{ESD} \right\}$ for fine-tuning our models. 
 For each of the datasets used, we generate mel-spectrograms of the speech data. 
 % \textcolor{red}{Given a set of mel-spectrograms extracted from the speech data, the task is to classify each spectrogram into one of four emotion categories: neutral, happy, sad, and angry.}
 Given a set of mel-spectrograms extracted from the speech data, the task is to classify each spectrogram into one of four emotion categories: neutral, happy, sad, and angry. Each spectrogram,  $\bm{x}_{i}^{d}$, where $x\in \mathbf{R}^{H \times W \times C}$, $d \in \left\{\textrm{RAVDESS}, \textrm{TESS}, \textrm{CREMA-D}, \textrm{ESD}, \textrm{MELD} \right\}$, and $i$ is the index of the datapoint, is passed through two pipelines (see Figure \ref{fig:ViT_model}, both ViT and BEiT encoders) to evaluate the performance of vision transformers for speech emotion recognition tasks. Here $H = 224, W = 224, C = 3$, represent the height, width, and the number of channels of the image respectively. 

The formulation of both of these pipelines remains the same with the only difference of using a pre-trained base ViT encoder (vit-base-patch16-224) for the first pipeline (ViT encoder as the transformer encoder in Figure \ref{fig:ViT_model}) whereas using a base BEiT encoder (microsoft/beit-base-patch16-224-pt22k-ft22k) for pipeline 2 (BEiT as the transformer encoder in Figure \ref{fig:ViT_model})\cite{dosovitskiy2020image, bao2021beit}. Each image $\bm{x}_{i}^{d}$ is first divided into patches, $x_{p} \in \mathbf{R}^{P \times P \times C}$, where P = 16 is the dimension of the image patch. So the output of the linear projection layer, $x' \in \mathbf{R}^{N \times (P^{2}C)}$, where $N$ is the number of patches. The patch and position embedding is then done using:
\begin{equation}
    z = \left[x_{[CLS]}, x' + \textrm{pos\_embed}\right] 
\end{equation}
\begin{equation}
    z_{norm} = LN(z)
\end{equation}
where LN(.) is the layer normalization layer and $\textrm{pos\_embed}$ is the position embedding added to each vector at the end of the linear projection layer. Then the values in the sequence are weighted through learnable matrices:  query (\textbf{q}), key (\textbf{k}), and value (\textbf{v}) to calculate self-attention given by the authors in \cite{vaswani2017attention, dosovitskiy2020image}:
\begin{equation}
    \left[\textbf{\textrm{q}}, \textbf{\textrm{k}}, \textbf{\textrm{v}}\right] = z_{norm}U_{qkv}
\end{equation}
where, $U_{qkv} \in \mathbf{R}^{D \times 3D_{h}}$ are learnable matrices. Then the self-attention is calculated as:
\begin{equation}
    SA(z_{norm}) = \textrm{softmax} \left(\frac{\textbf{\textrm{q}}\textbf{\textrm{k}}^{T}}{\sqrt{D_{h}}}\right)\textbf{\textrm{v}}
\end{equation}
So, the multihead attention, which is the multiple self attention operations in parallel heads can be expressed as \cite{vaswani2017attention, dosovitskiy2020image}:
\begin{dmath}
    MSA(z_{norm)} = \left[SA_{1}(z_{norm});SA_{2}(z_{norm});\\ \dots;SA_{k}(z_{norm})\right]U_{msa}
\end{dmath}
where, $U_{msa} \in \mathbf{R}^{k.D_{h} \times D}$, $D_{h}$ is the dimension of each head, $k$ is the number of attention heads, and $D$ is the dimension of the transformer model. The output of the transformers encoder is given by:
\begin{dmath}
   \widehat{y} =  (MSA(z_{norm}) + z )+ \\MLP(LN(MSA(z_{norm}) + z))
\end{dmath}
where MLP(.) is the multilayer perception.

% \begin{algorithm}
% \caption{Personalization model}
% \label{alg:example}
% \begin{algorithmic}[1] % The [1] adds line numbers
% \REQUIRE Input parameters $x$, $y$
% \ENSURE Output $z$
% \STATE Initialize $z \gets 0$
% \FOR{$i = 1$ to $n$}
%     \STATE $z \gets z + x[i] \cdot y[i]$
% \ENDFOR
% \RETURN $z$
% \end{algorithmic}
% \end{algorithm}

\begin{algorithm}
\caption{\textcolor{black}{Personalization Process for Speech Emotion Recognition}}
\label{alg:personalization}
\begin{algorithmic}[1]
\REQUIRE \textcolor{black}{Set of pre-defined questions $Q = \{q_1, q_2, ..., q_5\}$, Emotions $E = \{\text{neutral, happy, sad, angry}\}$}
\ENSURE \textcolor{black}{y: Accuracy, Precision, Recall, F1 Score, FLOPs, Average Inference Time}

\STATE \textcolor{black}{\textbf{Initialization:} Prepare robot for interaction.}
\FOR{\textcolor{black}{each emotion $e \in E$}}
    \FOR{\textcolor{black}{each question $q \in Q$}}
        \STATE \textcolor{black}{Person asks the robot question $q$.}
        \STATE \textcolor{black}{Robot responds to question $q$.}
        \STATE \textcolor{black}{Robot asks the same question $q$ back to the person.}
        \STATE \textcolor{black}{Person gives an open-ended reply to question $q$.}
    \ENDFOR
\ENDFOR

\STATE \textcolor{black}{\textbf{Data Collection:} Save all responses as audio files (.wav format).}

\STATE \textcolor{black}{\textbf{Data Preprocessing:}}
\STATE \textcolor{black}{Convert audio files into mel-spectrograms.}
\STATE \textcolor{black}{Perform stratified train-test split on the dataset.}

\STATE \textcolor{black}{\textbf{Model Fine-Tuning:}}
\STATE \textcolor{black}{Fine-tune the chosen model using the training dataset.}

\STATE \textcolor{black}{\textbf{Model Evaluation:}}
\STATE \textcolor{black}{Calculate metrics: Accuracy, Precision, Recall, F1 Score.}
\STATE \textcolor{black}{Compute FLOPs for one iteration.}
\STATE \textcolor{black}{Measure average inference time per sample.}
\STATE \textcolor{black}{\textbf{Output:} Return y}
\end{algorithmic}
\end{algorithm}

% \textcolor{red}{Ruchik, yanhan model 3 (ensemble) ka zikar kar sakte hein kya? directly results and discussion me ensemble model rakhna sahi he?}
% We use these vanilla-ViT and vanilla-BEiT models and fine-tune them on each of the benchmark datasets mentioned in Section \ref{datasets} using two different approaches which will be explained in Section \ref{results}. In addition to it, the personalized models for each participant has been explained in Section \ref{results} but fol

% The output of the transformer encoder is \cite{dosovitskiy2020image}:
% \begin{equation}
%    MHA(z) = \left[SA_{1}(z);SA_{2}(z); \dots; SA_{k}\right]U_{mha}
% \end{equation}
% where $U_{mha} \in \mathbf{R}^{k. D_{h}\times D}$, $k$ is the number of attention heads used, $D$ is the dimension of the model, $D_{h}$ is the dimension of each attention head, and $SA_{z}$ is the self attention calculated by:
% \begin{equation}
%     SA(z) = Av
% \end{equation}
% where, 
% \begin{equation}
% A = \textrm{softmax} \left(\frac{qk^{T}}{\sqrt{D_{h}}}\right)
% \end{equation}
% is the attention calculated over all the values in the sequence $v$ \cite{dosovitskiy2020image}, and $q,k,v$ are the query, key, and value matrices.

\section{Results and Discussion}\label{results}
We evaluate both the ViT and the BEiT pipelines in two ways:
\begin{itemize}
    \item \textbf{Approach 1: } In this approach, we train the individual ViT$_{d}$ and BEiT$_{d}$ models, where $d \in \left\{\textrm{RAVDESS}, \textrm{TESS}, \textrm{CREMA-D}, \textrm{ESD}, \textrm{MELD} \right\}$. We split each of the datasets, $(\mathcal{X}_{d}, \mathcal{Y}_{d})$ into $(\mathcal{X}_{d,train}, \mathcal{Y}_{d,train})$ and $(\mathcal{X}_{d,test}, \mathcal{Y}_{d,test})$. Then we train separate ViT$_{d}$ and BEiT$_{d}$ models, individually for each of these datasets. Since we have a four class classification problem of classifying the mel spectrograms into four primary emotions, we use cross entropy loss. 
    
    % The output logits of each of these models can be denoted by $c_{d}$ where $d \in \left\{\textrm{RAVDESS}, \textrm{TESS}, \textrm{CREMA-D}, \textrm{ESD}, \textrm{MELD} \right\}$ and $c_{d} \in \mathbf{R}^{1 \times 5}$.

     % \begin{equation}
     %     \mathcal{C} = \left(c_{RAVDESS}, c_{TESS},c_{CREMA-D},c_{ESD},c_{MELD}\right)
     % \end{equation}
     % where $c_{i} \in \mathcal{C}$ and $c_{i} \in \mathbf{R}^{1 \times 5}$
    % We first fine-tune the ViT and the BEiT models on individual datasets (see Figure \ref{fig:individual_datasets}). For $\bm{x}_{i}^{d} \in \mathcal{X}^{d}$, we randomly split $\mathcal{X}^{d}$ into $\mathcal{X}^{d}_{train}$ and $\mathcal{X}^{d}_{test}$ datasets. Then we fine-tune the ViT and the BEiT models individually on each dataset with a 5-fold cross-validation with cross entropy loss. 
    
\item \textbf{Approach 2: } In this we combine the datasets together:
\begin{equation}
    \mathcal{X}_{train, mix}, \mathcal{Y}_{train, mix} = \left( \bigcup_d \mathcal{X}_{d,train}, \bigcup_d \mathcal{Y}_{d,train} \right)
\end{equation}
and then fine-tune a ViT$_{mix}$ and a BEiT$_{mix}$ model on this mix training set $\mathcal{X}_{train, mix}, \mathcal{Y}_{train, mix}$. 
% For this approach, we combine the training sets of all datasets, and then fine tune our ViT and BEiT pipelines individually (see Figure \ref{fig:mix_model}). We use 5-fold cross-validation during our fine-tuning process and then test the models on $\mathcal{X}^{d}_{test}$. 
\end{itemize}
We perform full fine-tuning of our models on two A5000 GPUs, using K-Fold-Cross validation (5-fold-cross-validation in our case) with a constant learning rate of $2.00e-05$. Further, we evaluate the performance of both pipelines for both Approach 1 and 2 using accuracy, precision, recall, and f-1 scores. 

\begin{figure*}[h!]
    \centering
    \subfloat[ViT embeddings]{%
        \includegraphics[width=0.85\textwidth]{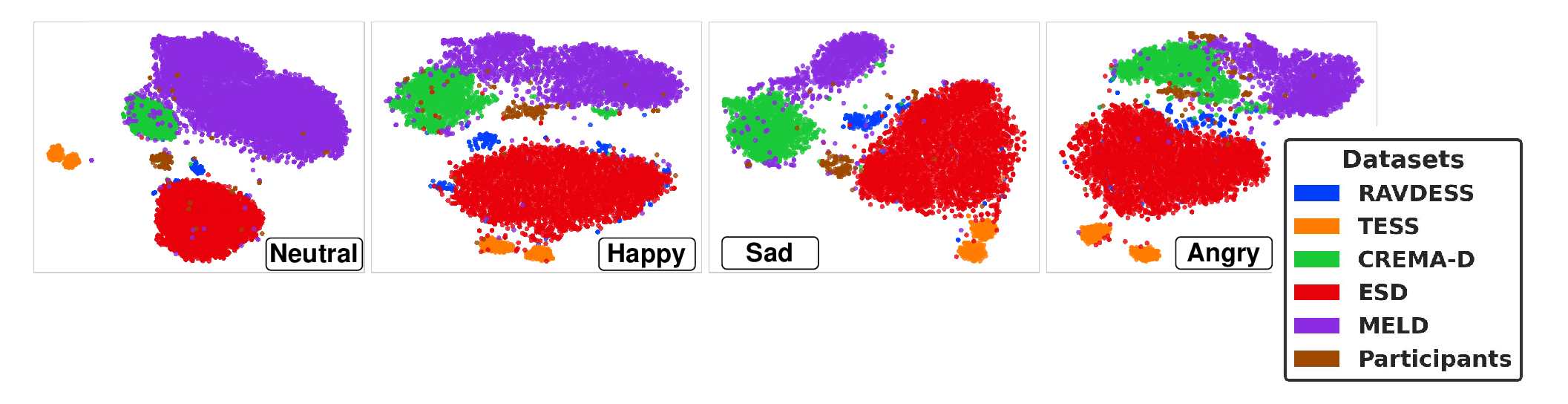}%
        \label{fig:tsne_vit}
    }
\vspace{-0.4cm}

     \subfloat[BEiT embeddings]{%
    \includegraphics[width=0.75\textwidth]{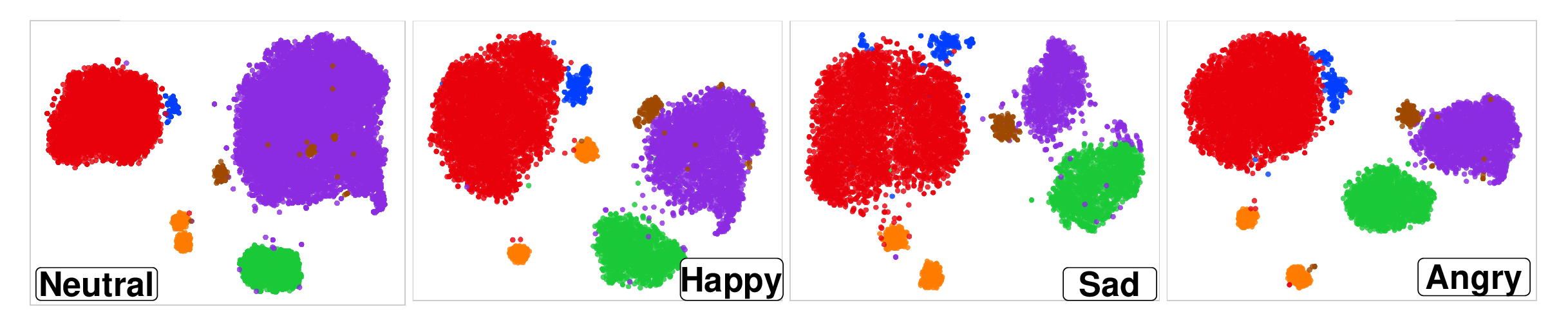}%
        \label{fig:tsne_beit}
    }
    \caption{\color{black}T-SNE plots of ViT and BEiT embeddings for each emotion of all datasets and our collected participants' data. The feature space of the emotional representations for the ViT and the BEiT models for each emotion is shown for all benchmark datasets as well as the participant data.}
    \label{fig:tsne_plots}
\end{figure*}

Table \ref{tab:performance_metrics} compares results of Approach 1 and Approach 2 for ViT (ViT$_{d}$ and ViT$_{mix}$), BEiT (BEiT$_{d}$ and BEiT$_{mix}$), OpenAI/Whisper-base (openai/whisper-base and openai/whisper-base$_{mix}$), ResNeT-50 (ResNeT-50$_{d}$ and ResNeT-50$_{mix}$) models. For the RAVDESS dataset, we currently achieve SOTA using the vanilla-ViT model, with the highest performance of 97.49\% accuracy as compared to the current SOTA, which has a classification accuracy of 86.70\% using multimodal data \cite{livingstone2018ryerson}. Vanilla-ViT model also outperforms  OpenAI/Whisper-base and ResNet-50 models. For the TESS dataset, we again achieve SOTA using vanilla-ViTs and vanilla-BEiTs, which is very similar to the ones obtained by the authors in \cite{akinpelu2024enhanced}, \textcolor{black} {openai/whisper-base model, and ResNet-50 model}. \textcolor{black}{Among our vision transformer based approaches}, the classification accuracy for the CREMA-D dataset was the highest for the mixed dataset approach (Approach 2) with vanilla-ViTs, which is better than the performance of comparable transformer architectures presented by the authors in \cite{ristea2022septr} and other non-transformer-based approaches \cite{ristea2021self,georgescu2023nonlinear}. \textcolor{black}{However, among all the approaches, openai/whisper-base performed the best (80.82\%) for the CREMA-D dataset when it was fine-tuned on only the CREMA-D training set}. For the ESD dataset, our peak classification accuracy (96.25\%) was obtained by a vanilla-BEiT model fine-tuned only on $(\mathcal{X}_{ESD,train}, \mathcal{Y}_{ESD,train})$, which is again comparable to the current SOTA (93.20\%) as presented by the authors in \cite{saliba2024layer}. \textcolor{black}{It also outperforms openai/whisper-base and ResNet-50 based approach we examined. } Since MELD dataset has numerous speakers, it covers a wide-range of speaker characteristics (see Figure \ref{fig:tsne_plots}). This can be see in the low classification accuracy of the MELD dataset from the Table \ref{tab:performance_metrics}. \textcolor{black}{Among our ViT and BEiT models,} we obtained peak accuracy when the BEiT model fine-tuned over $(\mathcal{X}_{train, mix}, \mathcal{Y}_{train, mix})$. However, our results with the MELD come close to the classification accuracies presented by the authors in \cite{ibrahim2024does}. \textcolor{black}{In addition to it, based on our experiments, we observed the highest classification accuracies for MELD$_{mix}$ with openai/whisper-base model.}

\begin{table*}[!t]
\arrayrulecolor{black}
\centering
\scriptsize                 % or \footnotesize
\setlength{\tabcolsep}{3pt} % tighten column spacing
\renewcommand{\arraystretch}{0.95} % slightly reduce row spacing
\caption{\textcolor{black}{Performance on five emotion datasets for Approach 1 (Unmixed) and Approach 2 (Mixed).
``--'' indicates unavailable metrics. M1--M5 are explained below. The learning rate used for all the models was 2.00e-5.}}
\label{tab:performance_metrics}

%----------------------------------------
%   A P P R O A C H   1 (UNMIXED)
%----------------------------------------
\textbf{Approach 1 (Unmixed Data)}
\vspace{0.5em}

\begin{tabular}{|c|cccc|cccc|cccc|cccc|cccc|}
\hline
    \rowcolor{black!0}
\multicolumn{1}{|c|}{} &
\multicolumn{4}{c|}{\textbf{RAVDESS}} &
\multicolumn{4}{c|}{\textbf{TESS}} &
\multicolumn{4}{c|}{\textbf{CREMA-D}} &
\multicolumn{4}{c|}{\textbf{ESD}} &
\multicolumn{4}{c|}{\textbf{MELD}} \\
\cline{2-21}
\rowcolor{black!0}
\textbf{ID} 
& \textbf{Acc} & \textbf{P} & \textbf{R} & \textbf{F1}
& \textbf{Acc} & \textbf{P} & \textbf{R} & \textbf{F1}
& \textbf{Acc} & \textbf{P} & \textbf{R} & \textbf{F1}
& \textbf{Acc} & \textbf{P} & \textbf{R} & \textbf{F1}
& \textbf{Acc} & \textbf{P} & \textbf{R} & \textbf{F1} \\
\hline
%--- Row: M1
    \rowcolor{black!0}
M1 & \textbf{97.49} & 0.9749 & 0.9749 & 0.9749
   & 100.0 & 1.0000 & 1.0000 & 1.0000
   & 72.06 & 0.7237 & 0.7206 & 0.7213
   & 95.84 & 0.9585 & 0.9584 & 0.9584
   & 49.83 & 0.4402 & 0.4983 & 0.4601 \\
%--- Row: M2
    \rowcolor{black!0}
M2 & 94.62 & 0.9486 & 0.9462 & 0.9463
   & 100.0 & 1.0000 & 1.0000 & 1.0000
   & 71.85 & 0.7200 & 0.7185 & 0.7173
   & 96.25 & 0.9626 & 0.9625 & 0.9625
   & 43.32 & 0.4304 & 0.4332 & 0.4317 \\
       \rowcolor{black!0}
%--- Row: M3
M3 & 84.40 & 0.8876 & 0.7905 & 0.8059
   & 100.0 & 1.0000 & 1.0000 & 1.0000
   & \textbf{80.82} & 0.8103 & 0.8036 & 0.8051
   & \textbf{97.14} & 0.9716 & 0.9714 & 0.9715
   &\textbf{55.97} & 0.4168 & 0.3822 & 0.3925 \\
       \rowcolor{black!0}
%--- Row: M4
M4 & 65.67 & 0.7002 & 0.6567 & 0.6651
   & 100.0 & 1.0000 & 1.0000 & 1.0000
   & 70.41 & 0.7026 & 0.7041 & 0.6999
   & 90.65 & 0.9081 & 0.9065 & 0.9067
   & 51.12 & 0.4533 & 0.5112 & 0.4747 \\
       \rowcolor{black!0}
%--- Row: M5
M5 & --    & --     & --     & --
   & --    & --     & --     & --
   & 80.60 & --     & --     & --
   & --    & --     & --     & --
   & 53.00 & --     & --     & -- \\
\hline
\end{tabular}

\vspace{1em}

%----------------------------------------
%   A P P R O A C H   2 (MIXED)
%----------------------------------------
\textbf{Approach 2 (Mixed Data)}
\vspace{0.5em}

\begin{tabular}{|c|cccc|cccc|cccc|cccc|cccc|}
\hline
    \rowcolor{black!0}
\multicolumn{1}{|c|}{} &
\multicolumn{4}{c|}{\textbf{RAVDESS\_mix}} &
\multicolumn{4}{c|}{\textbf{TESS\_mix}} &
\multicolumn{4}{c|}{\textbf{CREMA-D\_mix}} &
\multicolumn{4}{c|}{\textbf{ESD\_mix}} &
\multicolumn{4}{c|}{\textbf{MELD\_mix}} \\
\cline{2-21}
    \rowcolor{black!0}
\textbf{ID}
& \textbf{Acc} & \textbf{P} & \textbf{R} & \textbf{F1}
& \textbf{Acc} & \textbf{P} & \textbf{R} & \textbf{F1}
& \textbf{Acc} & \textbf{P} & \textbf{R} & \textbf{F1}
& \textbf{Acc} & \textbf{P} & \textbf{R} & \textbf{F1}
& \textbf{Acc} & \textbf{P} & \textbf{R} & \textbf{F1} \\
\hline
    \rowcolor{black!0}
%--- Row: M1
M1 & \textbf{95.70} & 0.9572 & 0.9570 & 0.9570
   & 100.0 & 1.0000 & 1.0000 & 1.0000
   & 74.51 & 0.7522 & 0.7451 & 0.7467
   & 95.13 & 0.9513 & 0.9513 & 0.9513
   & 49.48 & 0.4413 & 0.4948 & 0.4594 \\
%--- Row: M2
    \rowcolor{black!0}
M2 & 94.98 & 0.9498 & 0.9498 & 0.9497
   & 100.0 & 1.0000 & 1.0000 & 1.0000
   & 72.36 & 0.7281 & 0.7236 & 0.7217
   & 95.28 & 0.9533 & 0.9528 & 0.9528
   & 50.22 & 0.4480 & 0.5022 & 0.4638 \\
%--- Row: M3
    \rowcolor{black!0}
M3 & 72.59 & 0.7873 & 0.7152 & 0.7083
   & 100.0 & 1.0000 & 1.0000 & 1.0000
   & \textbf{80.00} & 0.8068 & 0.7994 & 0.7993
   & \textbf{96.07} & 0.9612 & 0.9607 & 0.9606
   & \textbf{56.70} & 0.4458 & 0.4021 & 0.4162 \\
%--- Row: M4
    \rowcolor{black!0}
M4 & 74.63 & 0.7517 & 0.7463 & 0.7474
   & 99.34 & 0.9936 & 0.9934 & 0.9934
   & 74.49 & 0.7425 & 0.7449 & 0.7406
   & 89.44 & 0.8959 & 0.8944 & 0.8946
   & 50.33 & 0.4747 & 0.5033 & 0.4861 \\
%--- Row: M5
    \rowcolor{black!0}
M5 & -- & -- & -- & --
   & -- & -- & -- & --
   & -- & -- & -- & --
   & -- & -- & -- & --
   & -- & -- & -- & -- \\
\hline
\end{tabular}

\vspace{0.5em}
\begin{minipage}{0.9\textwidth}
\scriptsize
\textcolor{black}{\textbf{Model References:}\\
M1: ViT (google/vit-base-patch16-224) \quad
M2: BEiT (microsoft/beit-base-patch16-224) \quad
M3: Whisper (openAI/whisper-base) \quad
M4: ResNet50 \quad
M5: Vesper}
\end{minipage}
\end{table*}
\vspace{-0.4cm}
% \begin{figure*}[h!]
%     \centering
%     \subfloat[Participant 1]{%
%         \includegraphics[width=0.32\textwidth]{cm_p1.eps}%
%         \label{fig:p1}
%     }
%      \subfloat[Participant 2]{%
%         \includegraphics[width=0.32\textwidth]{cm_p2.eps}%
%         \label{fig:p2}
%     }
%       \subfloat[Participant 3]{%
%         \includegraphics[width=0.32\textwidth]{cm_p3.eps}%
%         \label{fig:p3}
%     }
    
%        \subfloat[Participant 4]{%
%         \includegraphics[width=0.32\textwidth]{cm_p4.eps}%
%         \label{fig:p4}
%     }
%      \subfloat[Participant 5]{%
%         \includegraphics[width=0.32\textwidth]{cm_p5.eps}%
%         \label{fig:p5}
%     }
%      \subfloat[Participant 6]{%
%         \includegraphics[width=0.32\textwidth]{cm_p6.eps}%
%         \label{fig:p6}
%     }
%     \caption{Confusion matrices for participants' data classification using their best individual models.}
%     \label{fig:cm_participants}
% \end{figure*}
\FloatBarrier  % Forces placement of the previous floats before proceeding
\begin{table}[h!]
\arrayrulecolor{black}
\centering
\caption{Number of samples for training and test per participant. Here $i$ denotes the participant number. $i \in \left\{1,2,3,4,5,6\right\}$}
\resizebox{\columnwidth}{!}{  % Resize table to fit within the column width
\begin{tabular}{|c|c|c|c|c|c|c|}
\hline
 & Data & \textbf{Neutral} & \textbf{Happy} & \textbf{Sad} & \textbf{Angry} & \textbf{Total} \\ \hline
\multirow{2}{*}{Participant$_{i}$} & Train & 6 & 6 & 6 & 6 & 24\\ \cline{2-7}
                       & Test & 4 & 4 & 4 & 4 & 16 \\ \hline
\end{tabular}
}  % End of resize box
\label{tab:participant}
\end{table}
% \vspace{-0.1cm}
\textbf{Remark 1:}
For a conversational dataset like MELD, methods like meta-learning for few-shot learning, and parameter efficient fine-tuning (PEFT) methods can help learn natural emotions in speech in addition to acted ones \cite{finn2017model, lashkarashvili2024parameter} for better domain adaptation.
% Domain adaptation techniques may further improve accuracy on the MELD dataset by helping the model handle the diverse range of speakers and conversational dynamics, that differ from other SER datasets. Some state-of-the-art techniques like multi-stage fine-tuning, speaker normalization, and domain adversarial training may enable the model to learn emotion representations that are more robust to speaker variability and adaptable to real-world conversational contexts.

\vspace{-0.5cm}
\subsection{Human subjects' study}
\arrayrulecolor{black}
\begin{table*}[h!]
    \begin{center}
    \caption{\textcolor{black}{Performance Metrics for Each Participant and Model. \newline Model Mapping: Model 1 - Vanilla-ViT, Model 2 - Vanilla-BEiT, Model 3 - ViT$_{mix}$, Model 4 - BEiT$_{mix}$, Model 5 - ViT$_{ensemble}$, Model 6 - BEiT$_{ensemble}$, Model 7 - ViT$_{ensemble,d}$, Model 8 - BEiT$_{ensemble,d}$, Model 9 - openai/Whisper-base, Model 10 - openai/whisper-base$_{mix}$, Model 11 - resnet-50, Model 12 - resnet-50$_{mix}$.}}
    \resizebox{\textwidth}{!}{
    \begin{tabular}{|c|c|c|c|c|c|c|c|c|c|c|c|c|c|c|c|}
     \rowcolor{black!0}
        \hline
        \textbf{Participant} & \textbf{Metric} & \textbf{Model 1} & \textbf{Model 2} & \textbf{Model 3} & \textbf{Model 4} & \textbf{Model 5} & \textbf{Model 6} & \textbf{Model 7} & \textbf{Model 8} & \textbf{Model 9} & \textbf{Model 10} & \textbf{Model 11} & \textbf{Model 12} \\
        \hline
        \rowcolor{black!0}
        \multirow{4}{*}{1} 
            & Accuracy (\%) & 56.25 & 62.5 & 68.75 & \textbf{75.00} & 68.75 & 56.25 & 68.75 & \textbf{75.00} & 68.75 & 75.00 & 50.00 & 68.75 \\
                \rowcolor{black!0}
           1 & Precision & 0.5625 & 0.6458 & 0.7125 & 0.8 & 0.7738 & 0.5089 & 0.7946 & 0.8304 & 0.725 & 0.8166 & 0.40 & 0.725 \\
                \rowcolor{black!0}
            & Recall    & 0.5625 & 0.625 & 0.6875 & 0.75 & 0.6875 & 0.5625 & 0.6875 & 0.75 & 0.6875 & 0.75 & 0.5 & 0.6875 \\
                \rowcolor{black!0}
            & F1 Score  & 0.5486 & 0.6208 & 0.6935 & 0.75 & 0.6959 & 0.5284 & 0.79 & 0.7193 & 0.700 & 0.7166 & 0.4166 & 0.70 \\
        \hline
            \rowcolor{black!0}
        \multirow{4}{*}{2} 
            & Accuracy (\%) & 37.5 & 43.75 & \textbf{56.25} & 37.5 & 37.5 & 43.75 & 43.75 & \textbf{56.25} & 43.75 & 43.75 & 6.25 & 25.00 \\
                \rowcolor{black!0}
          2  & Precision & 0.2905 & 0.333 & 0.5833 & 0.3854 & 0.2708 & 0.25 & 0.35 & 0.6625 & 0.45 & 0.677 & 0.0416 & 0.175 \\
                \rowcolor{black!0}
            & Recall    & 0.375 & 0.4375 & 0.5625 & 0.375 & 0.375 & 0.4375 & 0.4375 & 0.5625 & 0.4375 & 0.4375 & 0.0625 & 0.25 \\
                \rowcolor{black!0}
            & F1 Score  & 0.3189 & 0.3631 & 0.5607 & 0.3512 & 0.3006 & 0.3154 & 0.3611 & 0.5486 & 0.4365 & 0.425 & 0.05 & 0.2055 \\
        \hline
            \rowcolor{black!0}
        \multirow{4}{*}{3} 
            & Accuracy (\%) & 43.75 & 43.75 & 25.00 & \textbf{56.00} & 50.00 & 43.75 & 50.00 & 43.75 & 25.00 & 18.75 & 25.00 & 31.25 \\
                \rowcolor{black!0}
           3 & Precision & 0.3571 & 0.5769 & 0.3125 & 0.7875 & 0.6111 & 0.4405 & 0.6417 & 0.475 & 0.196 & 0.125 & 0.22 & 0.2499 \\
                \rowcolor{black!0}
            & Recall    & 0.4375 & 0.4375 & 0.25 & 0.5625 & 0.5 & 0.4375 & 0.5 & 0.4375 & 0.25 & 0.1875 & 0.25 & 0.3125 \\
                \rowcolor{black!0}
            & F1 Score  & 0.3697 & 0.3843 & 0.1938 & 0.5304 & 0.4622 & 0.4161 & 0.469 & 0.3679 & 0.2123 & 0.1468 & 0.2197 & 0.275 \\
        \hline
            \rowcolor{black!0}
        \multirow{4}{*}{4} 
            & Accuracy (\%) & 50.00 & 50.00 & 50.00 & 37.5 & \textbf{75.00} & 62.5 & 56.25 & 62.5 & 37.5 & 50.00 & 25.00 & 50.00 \\
                \rowcolor{black!0}
           4 & Precision & 0.375 & 0.583 & 0.4583 & 0.3854 & 0.7875 & 0.7986 & 0.525 & 0.6667 & .2986 & 0.5821 & 0.076 & 0.4875 \\
                \rowcolor{black!0}
            & Recall    & 0.5 & 0.5 & 0.5 & 0.375 & 0.75 & 0.625 & 0.5625 & 0.625 & 0.375 & 0.5 & 0.25 & 0.5 \\
                \rowcolor{black!0}
            & F1 Score  & 0.4278 & 0.4393 & 0.4679 & 0.3512 & 0.7431 & 0.608 & 0.5214 & 0.6071 & 0.2996 & 0.4974 & 0.1176 & 0.479 \\
        \hline
            \rowcolor{black!0}
        \multirow{4}{*}{5} 
            & Accuracy (\%) & 37.5 & 25.00 & 37.5 & 37.5 & 43.75 & \textbf{50.00} & 31.25 & 43.75 & 37.5 & 25.00 & 25.00 & 37.5 \\
                \rowcolor{black!0}
           5 & Precision & 0.211 & 0.3527 & 0.3708 & 0.425 & 0.333 & 0.5833 & 0.375 & 0.4571 & 0.6375 & 0.1916 & 0.0625 & 0.333 \\
                \rowcolor{black!0}
            & Recall    & 0.375 & 0.25 & 0.375 & 0.375 & 0.4375 & 0.5 & 0.3125 & 0.4375 & 0.375 & 0.25 & 0.25 & 0.375 \\
                \rowcolor{black!0}
            & F1 Score  & 0.265 & 0.23236 & 0.37 & 0.3631 & 0.375 & 0.4631 & 0.3167 & 0.4141 & 0.3696 & 0.2166 & 0.1 & 0.29166 \\
        \hline
            \rowcolor{black!0}
        \multirow{4}{*}{6} 
            & Accuracy (\%) & 56.25 & 43.75 & 37.50 & 43.75 & \textbf{62.50} & 50.00 & 56.25 & 43.75 & 50.00 & 50.00 & 31.25 & 43.75 \\
                \rowcolor{black!0}
          6  & Precision & 0.6 & 0.4146 & 0.4015 & 0.3917 & 0.70 & 0.5 & 0.75 & 0.4208 & 0.5833 & 0.6071 & 0.1905 & 0.4687 \\
                \rowcolor{black!0}
            & Recall   & 0.5625 & 0.4375 & 0.375 & 0.4375 & 0.625 & 0.5 & 0.5625 & 0.4375 & 0.500 & 0.500 & 0.3125 & 0.4375 \\
                \rowcolor{black!0}
            & F1 Score  & 0.5754 & 0.4206 & 0.3381 & 0.3944 & 0.5972 & 0.4667 & 0.5916 & 0.4256 & 0.4714 & 0.4864 & 0.2364 & 0.4226 \\
        \hline
            \rowcolor{black!0}
        \multirow{4}{*}{7} 
            & Accuracy (\%) & 46.67 & 33.33 & \textbf{53.33} & 13.33 & 46.67 & 40.00 & 53.33 & 46.67 & 18.75 & 18.75 & 31.25 & 43.75 \\
                \rowcolor{black!0}
          7  & Precision & 0.44 & 0.422 & 0.4778 & 0.0667 & 0.5889 & 0.544 & 0.6267 & 0.4667 & 0.1548 & 0.1458 & 0.2019 & 0.4688 \\
                \rowcolor{black!0}
            & Recall    & 0.4667 & 0.33 & 0.5333 & 0.133 & 0.4667 & 0.400 & 0.533 & 0.4667 & 0.1875 & 0.1875 & 0.3125 & 0.4375 \\
                \rowcolor{black!0}
            & F1 Score  & 0.4487 & 0.3022 & 0.4857 & 0.0889 & 0.4610 & 0.3859 & 0.511 & 0.4222 & 0.1623 & 0.1625 & 0.201 & 0.4167 \\
        \hline
            \rowcolor{black!0}
           \multirow{4}{*}{8} 
            & Accuracy (\%) & 37.50 & 25.00 & 37.50 & 37.50 & 37.50 & 37.50 & \textbf{56.25} & 31.25 & 25.00 & 6.25 & 25.00 & 18.75 \\
                \rowcolor{black!0}
          8  & Precision & 0.4196 & 0.1500 & 0.3750 & 0.3155 & 0.4196 & 0.200 & 0.5792 & 0.2458 & 0.0769 & 0.0321 & 0.0625 & 0.206 \\
                \rowcolor{black!0}
            & Recall    & 0.3750 & 0.250 & 0.3750 & 0.3750 & 0.3750 & 0.3750 & 0.5625 & 0.3125 & 0.250 & 0.0625 & 0.2500 & 0.1875 \\
                \rowcolor{black!0}
            & F1 Score  & 0.3197 & 0.1825 & 0.3416 & 0.3292 & 0.3197 & 0.2540 & 0.565 & 0.2736 & 0.1176 & 0.0417 & 0.10 & 0.1806 \\
        \hline
            \rowcolor{black!0}
           \multirow{4}{*}{9} 
            & Accuracy (\%) & 43.75 & \textbf{56.25} & 25.00 & 31.25 & 43.75 & 37.50 & 31.25 & 37.50 & 31.25 & 37.50 & 18.75 & 37.50 \\
                \rowcolor{black!0}
           9 & Precision & 0.3792 & 0.4304 & 0.2167 & 0.1562 & 0.4107 & 0.5089 & 0.3125 & 0.3875 & 0.3083 & 0.3214 & 0.1056 & 0.4167 \\
                \rowcolor{black!0}
            & Recall    & 0.4375 & 0.5625 & 0.2500 & 0.3125 & 0.4375 & 0.3750 & 0.3125 & 0.3750 & 0.3125 & 0.3750 & 0.1875 & 0.3750 \\
                \rowcolor{black!0}
            & F1 Score  & 0.4042 & 0.4804 & 0.2306 & 0.2083 & 0.4205 & 0.3784 & 0.3054 & 0.3681 & 0.2944 & 0.3409 & 0.1325 & 0.3667 \\
        \hline
            \rowcolor{black!0}
           \multirow{4}{*}{10} 
            & Accuracy (\%) & 37.50 & 37.50 & 56.25 & 43.75 & 50.00 & \textbf{62.50} & 37.50 & 56.25 & 18.75 & 56.25 & 25.00 & 31.25 \\
                \rowcolor{black!0}
           10 & Precision & 0.300 & 0.2798 & 0.5458 & 0.4405 & 0.500 & 0.6875 & 0.3583 & 0.5833 & 0.1458 & 0.4446 & 0.0667 & 0.400 \\
                \rowcolor{black!0}
            & Recall    & 0.3750 & 0.3750 & 0.5625 & 0.4375 & 0.500 & 0.6250 & 0.3750 & 0.5625 & 0.1875 & 0.5625 & 0.25 & 0,3125 \\
                \rowcolor{black!0}
            & F1 Score  & 0.3056 & 0.3123 & 0.5506 & 0.4161 & 0.4921 & 0.6446 & 0.3631 & 0.5357 & 0.1548 & 0.4905 & 0.1053 & 0.333 \\
        \hline
            \rowcolor{black!0}
           \multirow{4}{*}{11} 
            & Accuracy (\%) & 60.00 & 46.67 & 60.00 & \textbf{73.33} & 53.33 & 40.00 & 60.00 & 66.67 & 60.00 & \textbf{73.33} & 40.00 & 40.00 \\
                \rowcolor{black!0}
          11  & Precision & 0.6457 & .2519 & 0.7852 & 0.7467 & 0.5733 & 0.3556 & 0.5778 & 0.7968 & 0.6250 & 0.7875 & 0.2067 & 0.5378 \\
                \rowcolor{black!0}
            & Recall    & 0.600 & 0.4667 & 0.600 & 0.7333 & 0.533 & 0.400 & 0.6 & 0.6667 & 0.5833 & 0.7292 & 0.400 & 0.400 \\
                \rowcolor{black!0}
            & F1 Score & 0.5606 & 0.3241 & 0.5708 & 0.7304 & 0.4590 & 0.3111 & 0.5511 & 0.6139 & 0.5893 & 0.7411 & 0.2667 & 0.4394 \\
        \hline
            \rowcolor{black!0}
           \multirow{4}{*}{12} 
            & Accuracy (\%) & 37.50 & 43.75 & 50.00 & 31.25 & \textbf{62.50} & 50.00 & \textbf{62.50} & 43.75 & 31.25 & 37.50 & 31.25 & 37.50 \\
                \rowcolor{black!0}
          12  & Precision & 0.433 & 0.6458 & 0.6071 & 0.2979 & 0.7778 & 0.4970 & 0.6417 & 0.3611 & 0.3006 & 0.3083 & 0.3167 & 0.4437 \\
                \rowcolor{black!0}
            & Recall    & 0.3750 & 0.4375 & 0.500 & 0.3125 & 0.6250 & 0.500 & 0.6250 & 0.4375 & 0.3125 & 0.3750 & 0.3125 & 0.3750 \\
                \rowcolor{black!0}
            & F1 Score  & 0.3265 & 0.4446 & 0.4697 & 0.2956 & 0.6300 & 0.4705 & 0.6290 & 0.3681 & 0.2963 & 0.3361 & 0.2053 & 0.3361 \\
        \hline
    \end{tabular}}
    \label{tab:model_performance}
      \end{center}
\end{table*}
% \vspace{-0.2cm}
We evaluated our speech emotion recognition in a pseudo-naturalistic human-robot interaction scenario using our fine-tuned ViTs and BEiTs. Since each participant asked five questions to the robot and responded to those five questions asked by the robot, we have 40 audio clips from each participant. We divided them into train and test datasets such that two sets of questions and answers each participant gave were separated for the test set. So, each participant had three questions and answer sets for train data. Each of those questions and answers was spoken in a way that depicts each of the four primary emotions of the individual. The split of the train and test data for each participant is shown in Table \ref{tab:participant}. Once the audio has been recorded from the participants, we convert the WAV files into spectrograms as shown in Figure \ref{fig:pipeline}. 

As described in Section \ref{data_acquisition}, each question-answer set was spoken in the four primary emotions. Hence, each participant had six audio clips for each emotion for the train set and four for the test set. Owing to the performance of Vision transformers-based approaches from Table \ref{tab:performance_metrics}, we used similar approaches to evaluate the use of vision transformers for speech emotion recognition in pseudo-naturalistic human-robot interaction. 
\begin{itemize}
    \item \textbf{Model 1 and 2- Vanilla-ViT and BEiT: } Each individual's data is converted to mel-spectrograms, and then vanilla-ViT and BEiT models are fine-tuned. 
    \item \textbf{Model 2 and 3- ViT$_{mix}$ and BEiT$_{mix}$: } The fine-tuned models from Approach 2 are fine-tuned on the participants' mel-spectrograms.
    \item \textbf{Model 3 and 4-  ViT$_{ensemble}$ and BEiT$_{ensemble}$: } We use five vanilla-ViTs and five vanilla-BEiTs and average the logits. If the output of each ViT$_{i}$, where $i=\left\{1,2,3,4,5\right\}$ and of each BEiT$_{i}$ are $c_{i,vit}$ and $c_{i,BEiT}$ respectively, then the ensemble of models is:
    \begin{equation}
        \textrm{ViT}_{ensemble} = \frac{1}{5}\sum_{i=1}^{5} c_{i,ViT} 
    \end{equation}
    \vspace{-0.2cm}
    \begin{equation}
        \textrm{BEiT}_{ensemble} = \frac{1}{5}\sum_{i=1}^{5} c_{i,BEiT} 
    \end{equation}
    \item \textbf{Model 5 and 6-  ViT$_{ensemble,d}$ and BEiT$_{ensemble,d}$: } In this approach, we use the ViT$_{d}$ and BEiT$_{d}$ models trained in Approach 1 on each of the benchmark datasets. So the ensemble works as follows:
      \begin{equation}
        \textrm{ViT}_{ensemble,d} = \frac{1}{5}\sum_{d}c_{ViT_{d}} 
    \end{equation}
    \begin{equation}
        \textrm{BEiT}_{ensemble,d} = \frac{1}{5}\sum_{d} c_{BEiT_{d}} 
    \end{equation}
\end{itemize}
\vspace{-0.2cm}
% \begin{table}[ht]
%     \centering
%     \caption{\color{black}Model complexity and inference times}
%     \begin{tabular}{lcc}
%         \hline
%                 \textbf{\color{black}Model} & \textbf{\color{black}FLOPs} & \textbf{\color{black}\shortstack{Inference Time \\ (ms/sample)}} \\
%         \hline
%         \color{black} Vanilla-ViT (Model 1) & 16.87 GMac & 0.516\\
%         \color{black}Vanilla-BEiT (Model 2) & 17.59 GMac & 3.315 \\
%         \color{black} ViT$_{mix}$ (Model 3) & 16.87 GMac & 0.4684\\
%         \color{black} BEiT$_{mix}$ (Model 4) & 17.59 GMac & 3.339\\
%         \color{black} ViT$_{ensemble}$ (Model 5) & 84.34 GMac & 2.418\\
%         \color{black} BEiT$_{ensemble}$ (Model 6) & 87.94 GMac & 16.6026\\
%          \color{black} ViT$_{ensemble,d}$ (Model 5) & 84.34 GMac & 2.188\\
%         \color{black} BEiT$_{ensemble,d}$ (Model 6) & 87.94 GMac & 13.581\\
%        \color{black} \color{black}OpenAI/Whisper-base (Model 7) & 30.11 GMac & 197.141 \\
%         \color{black} \color{black}OpenAI/Whisper-base$_{mix}$ (Model 8)& 30.11 GMac & 197.141 \\
%        \color{black}ResNet-50$_{mix}$ (Model 9) & 4.13 GMac & \color{black}0.517 \\
%         \color{black}ResNet-50$_{mix}$ (Model 10) & 4.13 GMac & \color{black}0.508 \\
%         \hline
%     \end{tabular}
%     \label{tab:model_comparison}
% \end{table}
\begin{table}[ht]
    \centering
    \caption{\color{black}Model complexity and inference times}
    \begin{tabular}{lcc}
        \hline
        \textbf{\color{black}Model} & \textbf{\color{black}FLOPs} & \textbf{\color{black}\shortstack{Inference Time \\ (ms/sample)}} \\
        \hline
        \color{black} Vanilla-ViT (Model 1) & \color{black}16.87 GMac & \color{black}0.516\\
        \color{black}Vanilla-BEiT (Model 2) & \color{black}17.59 GMac & \color{black}3.315 \\
        \color{black} ViT$_{mix}$ (Model 3) & \color{black}16.87 GMac & \color{black}0.4684\\
        \color{black} BEiT$_{mix}$ (Model 4) & \color{black}17.59 GMac & \color{black}3.339\\
        \color{black} ViT$_{ensemble}$ (Model 5) & \color{black}84.34 GMac & \color{black}2.418\\
        \color{black} BEiT$_{ensemble}$ (Model 6) & \color{black}87.94 GMac & \color{black}16.6026\\
        \color{black} ViT$_{ensemble,d}$ (Model 7) & \color{black}84.34 GMac & \color{black}2.188\\
        \color{black} BEiT$_{ensemble,d}$ (Model 8) & \color{black}87.94 GMac & \color{black}13.581\\
        \color{black}OpenAI/Whisper-base (Model 9) & \color{black}30.11 GMac & \color{black}197.141 \\
        \color{black}OpenAI/Whisper-base$_{mix}$ (Model 10)& \color{black}30.11 GMac & \color{black}197.141 \\
        \color{black}ResNet-50$_{mix}$ (Model 11) & \color{black}4.13 GMac & \color{black}0.517 \\
        \color{black}ResNet-50$_{mix}$ (Model 12) & \color{black}4.13 GMac & \color{black}0.508 \\
        \hline
    \end{tabular}
    \label{tab:model_comparison}
\end{table}
\textcolor{black}{Table IV shows the model performance of all the above proposed models. It becomes evident that the best performance is obtained when we use ViT or BEiT based approaches as compared to OpenAi/Whisper-base and ResNet-50. As can be seen from Figure \ref{fig:tsne_vit} and \ref{fig:tsne_beit}, the participant data has an overlap in the feature space of the datasets used in this paper. The overlap between the speech characteristics of speakers from these benchmark datasets and the participants for our human-robot interaction study helped better classify speech emotion compared to vanilla ViTs or vanilla-BEiTs. This contributes to the participants having better classification accuracies for the mix models and the ViT$_{ensemble,d}$/BEiT$_{ensemble,d}$ (see Table IV) for participants 1, 2, 3, 7, 8, 11, and 12. For some participants, the ensemble models (ViT$_{ensemble}$ and BEiT$_{ensemble}$) worked better since their speech characteristics didn't exactly overlap with the benchmark datasets used in this paper. For both native and non-native English speakers, ViT and BEiT based models performed better than other models compared. }
\textcolor{black}{For time complexity and inference times of our models, we analysed Floating Point Operations (FLOPs) and also recorded the average time it takes each of our models to classify one input test sample. As can be seen from Table \ref{tab:model_performance}, all of the participants had the best classification accuracies with either a ViT or BEiT based model except for participant 11, who had the same accuracy for the openai/whisper-base model too. However, the inference time for the openai/whisper-base was significantly higher (197.141 ms/sample) than the BEiT$_{mix}$ model (3.339 ms/sample).} Note that, real-time deployment of SER systems for HRI also depends on the system-specific requirements.
% should also consider memory footprint, processing power and energy constraints.\\

% The confusion matrix for each of the participants' best performing models have been shown in Figure \ref{fig:cm_participants}.
% \begin{table}[ht]
% \centering
% \caption{Table with subrows from the second column starting from row 2}
% \resizebox{\columnwidth}{!}{  % Resize table to fit within the column width
% \begin{tabular}{|c|c|c|c|c|c|}
% \hline
% Participants & Data & \textbf{Neutral} & \textbf{Happy} & \textbf{Sad} & \textbf{Angry} \\ \hline
% \multirow{2}{*}{Participant$_{i}$ & Train & 6 & 6 & 6 & 6 \\ \cline{2-6}
%                        & Test & 4 & 4 & 4 & 4 \\ \hline
% \multirow{2}{*}{Participant 2} & Train & 6 & 6 & 6 & 6 \\ \cline{2-6}
%                        & Test & 4 & 4 & 4 & 4\\ \hline
% \multirow{2}{*}{Participant 3} & Train & 6 & 6 & 6 & 6 \\ \cline{2-6}
%                        & Test & 4 & 4 & 4 & 4 \\ \hline
% \multirow{2}{*}{Participant 4} & Train & 6 & 6 & 6 & 6 \\ \cline{2-6}
%                        & Test & 4 & 4 & 4 & 4 \\ \hline
% \multirow{2}{*}{Participant 5} & Train & 6 & 6 & 6 & 6 \\ \cline{2-6}
%                        & Test & 4 & 4 & 4 & 4 \\ \hline
% \multirow{2}{*}{Participant 6} & Train & 6 & 6 & 6 & 6 \\ \cline{2-6}
%                        & Test & 4 & 4 & 4 & 4 \\ \hline
% \end{tabular}
% }  % End of resizebox
% \end{table}
% \section{Future Work}\label{future_work}
\vspace{-0.3cm}
\section{Ethics Statement}
Since this paper includes a human subjects' study, we took consent of all the participants on a consent form approved by the Institute Review Board (IRB Number: 18.0726). The participants had the opportunity to discontinue at any point of the study if they wanted to. 
\section{Conclusion and Future Works}\label{conclusion}
In this work, we address the gap in speech emotion recognition for pseudo-naturalistic and personalized verbal HRI. We evaluate the use of vision transformer based models for identifying four primary emotions: neutral, happy, sad, and angry from the speech characteristics of our participants' data. We do this by first fine-tuning the vision transformer-based models on benchmark datasets. We then use these fine-tuned models to fine-tune them again on participants' speech data and/or perform ensembling of these models. This helps us choose the best model for each participant, hence contributing towards understanding the emotional speech characteristics of each individual instead of proposing a group model. In addition to creating these personalized speech emotion recognition models, we also evaluate vanilla-ViT and vanilla-BEiTs on benchmark datasets like RAVDESS, TESS, CREMA-D, ESD, and MELD. We observed SOTA performances on some of these benchmark datasets.
% Furthermore, we also tested the use of fine-tuned ViT and BEiT models on a mix of these datasets and found that this fine-tuning on benchmark datasets boosted the classification accuracies for SER on our participants' data. For those participants that were outliers when visualized through t-sne plots (see Figure \ref{fig:tsne_plots}), we used ensemble methods to boost the classification accuracies. These two methods proved to be crucial in classifying SER in pseudo-naturalistic HRI. 
\vspace{-0.1cm}

In the future, we would like to recruit more human participants and collect data across different populations, including both neurotypical and neurodivergent populations. We would also like to examine multiple data modalities and examine how speech emotion correlates to modalities such as facial videos and physiological signals. 
\textcolor{black}{In addition to this, we would like to examine emotions on a more continuous scale, in terms of valence and arousal. This would help capture more subtle and complex emotions as compared to using only four discrete emotions, which is typically the case in human-human interactions.
Furthermore, we would also like to examine Few Shot Learning approaches for SER for datasets like MELD that have a large number of speakers \cite{finn2017model, snell2017prototypical}. This might help us generalize well for MELD since the current classification accuracies in the literature are comparatively lower as compared to other datasets.}
\vspace{-0.35cm}
\bibliographystyle{IEEEtran}
\bibliography{references}

\begin{thebibliography}{10}
\providecommand{\url}[1]{#1}
\csname url@samestyle\endcsname
\providecommand{\newblock}{\relax}
\providecommand{\bibinfo}[2]{#2}
\providecommand{\BIBentrySTDinterwordspacing}{\spaceskip=0pt\relax}
\providecommand{\BIBentryALTinterwordstretchfactor}{4}
\providecommand{\BIBentryALTinterwordspacing}{\spaceskip=\fontdimen2\font plus
\BIBentryALTinterwordstretchfactor\fontdimen3\font minus \fontdimen4\font\relax}
\providecommand{\BIBforeignlanguage}[2]{{%
\expandafter\ifx\csname l@#1\endcsname\relax
\typeout{** WARNING: IEEEtran.bst: No hyphenation pattern has been}%
\typeout{** loaded for the language `#1'. Using the pattern for}%
\typeout{** the default language instead.}%
\else
\language=\csname l@#1\endcsname
\fi
#2}}
\providecommand{\BIBdecl}{\relax}
\BIBdecl

\bibitem{johanson2021improving}
D.~L. Johanson, H.~S. Ahn, and E.~Broadbent, ``Improving interactions with healthcare robots: a review of communication behaviours in social and healthcare contexts,'' \emph{International Journal of Social Robotics}, vol.~13, no.~8, pp. 1835--1850, 2021.

\bibitem{Mishra2022}
R.~Mishra, ``Towards adaptive and personalized robotic therapy for children with autism spectrum disorder,'' in \emph{2022 10th International Conference on Affective Computing and Intelligent Interaction Workshops and Demos (ACIIW)}.\hskip 1em plus 0.5em minus 0.4em\relax IEEE, 2022, pp. 1--5.

\bibitem{Mishra2023}
R.~Mishra and K.~C. Welch, ``Towards forecasting engagement in children with autism spectrum disorder using social robots and deep learning,'' in \emph{SoutheastCon 2023}.\hskip 1em plus 0.5em minus 0.4em\relax IEEE, 2023, pp. 838--843.

\bibitem{nakanishi2020continuous}
J.~Nakanishi, I.~Kuramoto, J.~Baba, K.~Ogawa, Y.~Yoshikawa, and H.~Ishiguro, ``Continuous hospitality with social robots at a hotel,'' \emph{SN Applied Sciences}, vol.~2, pp. 1--13, 2020.

\bibitem{kirby2010affective}
R.~Kirby, J.~Forlizzi, and R.~Simmons, ``Affective social robots,'' \emph{Robotics and Autonomous Systems}, vol.~58, no.~3, pp. 322--332, 2010.

\bibitem{9384222}
M.~Pham, H.~M. Do, Z.~Su, A.~Bishop, and W.~Sheng, ``Negative emotion management using a smart shirt and a robot assistant,'' \emph{IEEE Robotics and Automation Letters}, vol.~6, no.~2, pp. 4040--4047, April 2021.

\bibitem{9508862}
K.~Maehama, J.~Even, C.~T. Ishi, and T.~Kanda, ``Enabling robots to distinguish between aggressive and joking attitudes,'' \emph{IEEE Robotics and Automation Letters}, vol.~6, no.~4, pp. 8037--8044, 2021.

\bibitem{Ibrahim2024}
A.~Ibrahim, S.~Shehata, A.~Kulkarni, M.~Mohamed, and M.~Abdul-Mageed, ``What does it take to generalize ser model across datasets? a comprehensive benchmark,'' \emph{arXiv preprint arXiv:2406.09933v1 [cs.SD]}, 2024.

\bibitem{Pepino2021}
L.~Pepino, P.~Riera, and L.~Ferrer, ``Emotion recognition from speech using wav2vec 2.0 embeddings,'' \emph{arXiv preprint arXiv:2104.03502v1 [cs.SD]}, 2021.

\bibitem{Bott2024}
T.~Bott, F.~Lux, and N.~T. Vu, ``Controlling emotion in text-to-speech with natural language prompts,'' \emph{arXiv preprint arXiv:2406.06406v2 [cs.CL]}, 2024.

\bibitem{Chang2022}
K.-W. Chang, W.-C. Tseng, S.-W. Li, and H.-y. Lee, ``Speechprompt: An exploration of prompt tuning on generative spoken language model for speech processing tasks,'' \emph{arXiv preprint arXiv:2203.16773v3 [eess.AS]}, 2022.

\bibitem{zhou2013speech}
P.~Zhou, X.-P. Li, J.~Li, and X.-X. Jing, ``Speech emotion recognition based on mixed mfcc,'' in \emph{Applied Mechanics and Mechanical Engineering III}.\hskip 1em plus 0.5em minus 0.4em\relax Trans Tech Publ, 2013.

\bibitem{lalitha2015speech}
S.~Lalitha, A.~Mudupu, B.~Nandyala, and R.~Munagala, ``Speech emotion recognition using dwt,'' in \emph{2015 IEEE International Conference on Computational Intelligence and Computing Research (ICCIC)}.\hskip 1em plus 0.5em minus 0.4em\relax IEEE, 2015, pp. 1--4.

\bibitem{rao2013emotion}
K.~S. Rao, S.~G. Koolagudi, and R.~R. Vempada, ``Emotion recognition from speech using global and local prosodic features,'' \emph{International Journal of Speech Technology}, 2013.

\bibitem{han2014speech}
K.~Han, D.~Yu, and I.~Tashev, ``Speech emotion recognition using deep neural network and extreme learning machine,'' in \emph{Proc. Interspeech}, 2014.

\bibitem{lee2015high}
J.~Lee and I.~Tashev, ``High-level feature representation using recurrent neural network for speech emotion recognition,'' in \emph{Proc. Interspeech}, 2015.

\bibitem{satt2017efficient}
A.~Satt, S.~Rozenberg, and R.~Hoory, ``Efficient emotion recognition from speech using deep learning on spectrograms,'' in \emph{Proc. Interspeech}, 2017.

\bibitem{khasgiwala2021vision}
Y.~Khasgiwala and J.~Tailor, ``Vision transformer for music genre classification using mel-frequency cepstrum coefficient,'' in \emph{2021 IEEE 4th International Conference on Computing, Power and Communication Technologies (GUCON)}.\hskip 1em plus 0.5em minus 0.4em\relax IEEE, 2021, pp. 1--5.

\bibitem{akinpelu2024enhanced}
S.~Akinpelu, S.~Viriri, and A.~Adegun, ``An enhanced speech emotion recognition using vision transformer,'' \emph{Scientific Reports}, vol.~14, no.~1, p. 13126, 2024.

\bibitem{lakomkin2018robustness}
E.~Lakomkin, M.~A. Zamani, C.~Weber, S.~Magg, and S.~Wermter, ``On the robustness of speech emotion recognition for human-robot interaction with deep neural networks,'' in \emph{2018 IEEE/RSJ International Conference on Intelligent Robots and Systems (IROS)}.\hskip 1em plus 0.5em minus 0.4em\relax IEEE, 2018, pp. 854--860.

\bibitem{mishra2024speech}
S.~Mishra, N.~Bhatnagar, P.~P, and S.~T.~R, ``Speech emotion recognition and classification using hybrid deep cnn and bilstm model,'' \emph{Multimedia Tools and Applications}, vol.~83, no.~13, pp. 37\,603--37\,620, 2024.

\bibitem{chen2020two}
L.~Chen, W.~Su, Y.~Feng, M.~Wu, J.~She, and K.~Hirota, ``Two-layer fuzzy multiple random forest for speech emotion recognition in human-robot interaction,'' \emph{Information Sciences}, vol. 509, pp. 150--163, 2020.

\bibitem{vaswani2017attention}
A.~Vaswani, ``Attention is all you need,'' \emph{arXiv preprint arXiv:1706.03762}, 2017.

\bibitem{luna2021proposal}
C.~Luna-Jim{\'e}nez, R.~Kleinlein, D.~Griol, Z.~Callejas, J.~M. Montero, and F.~Fern{\'a}ndez-Mart{\'\i}nez, ``A proposal for multimodal emotion recognition using aural transformers and action units on ravdess dataset,'' \emph{Applied Sciences}, vol.~12, no.~1, p. 327, 2021.

\bibitem{livingstone2018ryerson}
S.~R. Livingstone and F.~A. Russo, ``The ryerson audio-visual database of emotional speech and song (ravdess): A dynamic, multimodal set of facial and vocal expressions in north american english,'' \emph{PloS one}, vol.~13, no.~5, p. e0196391, 2018.

\bibitem{ibrahim2024does}
A.~Ibrahim, S.~Shehata, A.~Kulkarni, M.~Mohamed, and M.~Abdul-Mageed, ``What does it take to generalize ser model across datasets? a comprehensive benchmark,'' \emph{arXiv preprint arXiv:2406.09933}, 2024.

\bibitem{chumachenko2022self}
K.~Chumachenko, A.~Iosifidis, and M.~Gabbouj, ``Self-attention fusion for audiovisual emotion recognition with incomplete data,'' in \emph{2022 26th International Conference on Pattern Recognition (ICPR)}.\hskip 1em plus 0.5em minus 0.4em\relax IEEE, 2022, pp. 2822--2828.

\bibitem{dupuis2010tess}
K.~Dupuis and M.~K. Pichora-Fuller, ``Toronto emotional speech set (tess)-younger talker happy,'' University of Toronto, Psychology Department, Tech. Rep., 2010.

\bibitem{cao2014crema}
H.~Cao, D.~G. Cooper, M.~K. Keutmann, R.~C. Gur, A.~Nenkova, and R.~Verma, ``Crema-d: Crowd-sourced emotional multimodal actors dataset,'' \emph{IEEE transactions on affective computing}, vol.~5, no.~4, pp. 377--390, 2014.

\bibitem{chen2024vesper}
W.~Chen, X.~Xing, P.~Chen, and X.~Xu, ``Vesper: A compact and effective pretrained model for speech emotion recognition,'' \emph{IEEE Transactions on Affective Computing}, 2024.

\bibitem{saliba2024layer}
A.~Saliba, Y.~Li, R.~Sanabria, and C.~Lai, ``Layer-wise analysis of self-supervised acoustic word embeddings: A study on speech emotion recognition,'' \emph{arXiv preprint arXiv:2402.02617}, 2024.

\bibitem{liu2025personalized}
J.~Liu, M.~C. Ang, J.~K. Chaw, K.~W. Ng, and A.-L. Kor, ``Personalized emotion analysis based on fuzzy multi-modal transformer model,'' \emph{Applied Intelligence}, vol.~55, no.~3, p. 227, 2025.

\bibitem{ong2024maxmvit}
K.~L. Ong, C.~P. Lee, H.~S. Lim, K.~M. Lim, and A.~Alqahtani, ``Maxmvit-mlp: Multiaxis and multiscale vision transformers fusion network for speech emotion recognition,'' \emph{IEEE Access}, 2024.

\bibitem{mishra2023social}
R.~Mishra and K.~C. Welch, ``Social impressions of the nao robot and its impact on physiology,'' in \emph{2023 11th International Conference on Affective Computing and Intelligent Interaction Workshops and Demos (ACIIW)}.\hskip 1em plus 0.5em minus 0.4em\relax IEEE, 2023, pp. 1--8.

\bibitem{ancilin2021improved}
J.~Ancilin and A.~Milton, ``Improved speech emotion recognition with mel frequency magnitude coefficient,'' \emph{Applied Acoustics}, vol. 179, p. 108046, 2021.

\bibitem{abdul2022mel}
Z.~K. Abdul and A.~K. Al-Talabani, ``Mel frequency cepstral coefficient and its applications: A review,'' \emph{IEEE Access}, vol.~10, pp. 122\,136--122\,158, 2022.

\bibitem{zhang2019audio}
B.~Zhang, J.~Leitner, and S.~Thornton, ``Audio recognition using mel spectrograms and convolution neural networks,'' \emph{Noiselab University of California: San Diego, CA, USA}, 2019.

\bibitem{zhou2021seen}
K.~Zhou, B.~Sisman, R.~Liu, and H.~Li, ``Seen and unseen emotional style transfer for voice conversion with a new emotional speech dataset,'' in \emph{ICASSP 2021-2021 IEEE International Conference on Acoustics, Speech and Signal Processing (ICASSP)}.\hskip 1em plus 0.5em minus 0.4em\relax IEEE, 2021, pp. 920--924.

\bibitem{poria2018meld}
S.~Poria, D.~Hazarika, N.~Majumder, G.~Naik, E.~Cambria, and R.~Mihalcea, ``Meld: A multimodal multi-party dataset for emotion recognition in conversations,'' arXiv preprint arXiv:1810.02508, Tech. Rep., 2018.

\bibitem{dosovitskiy2020image}
A.~Dosovitskiy, L.~Beyer, A.~Kolesnikov, D.~Weissenborn, X.~Zhai, T.~Unterthiner, M.~Dehghani, M.~Minderer, G.~Heigold, S.~Gelly \emph{et~al.}, ``An image is worth 16x16 words: Transformers for image recognition at scale,'' \emph{arXiv preprint arXiv:2010.11929}, 2020.

\bibitem{bao2021beit}
H.~Bao, L.~Dong, S.~Piao, and F.~Wei, ``Beit: Bert pre-training of image transformers,'' \emph{arXiv preprint arXiv:2106.08254}, 2021.

\bibitem{ristea2022septr}
N.-C. Ristea, R.~T. Ionescu, and F.~S. Khan, ``Septr: Separable transformer for audio spectrogram processing,'' \emph{arXiv preprint arXiv:2203.09581}, 2022.

\bibitem{ristea2021self}
N.-C. Ristea and R.~T. Ionescu, ``Self-paced ensemble learning for speech and audio classification,'' \emph{arXiv preprint arXiv:2103.11988}, 2021.

\bibitem{georgescu2023nonlinear}
M.-I. Georgescu, R.~T. Ionescu, N.-C. Ristea, and N.~Sebe, ``Nonlinear neurons with human-like apical dendrite activations,'' \emph{Applied Intelligence}, vol.~53, no.~21, pp. 25\,984--26\,007, 2023.

\bibitem{finn2017model}
C.~Finn, P.~Abbeel, and S.~Levine, ``Model-agnostic meta-learning for fast adaptation of deep networks,'' in \emph{International conference on machine learning}.\hskip 1em plus 0.5em minus 0.4em\relax PMLR, 2017, pp. 1126--1135.

\bibitem{lashkarashvili2024parameter}
N.~Lashkarashvili, W.~Wu, G.~Sun, and P.~C. Woodland, ``Parameter efficient finetuning for speech emotion recognition and domain adaptation,'' in \emph{ICASSP 2024-2024 IEEE International Conference on Acoustics, Speech and Signal Processing (ICASSP)}.\hskip 1em plus 0.5em minus 0.4em\relax IEEE, 2024, pp. 10\,986--10\,990.

\bibitem{snell2017prototypical}
J.~Snell, K.~Swersky, and R.~Zemel, ``Prototypical networks for few-shot learning,'' \emph{Advances in neural information processing systems}, vol.~30, 2017.

\end{thebibliography}

\end{document}